\documentclass[prl,aps,twocolumn,showpacs,amssymb]{revtex4}
\usepackage{graphicx}
\usepackage{bm}

\begin{document}
\def\be{\begin{equation}}
\def\ee{\end{equation}}
\def\bte{\begin{eqnarray}}
\def\ete{\end{eqnarray}}
\def\bez{\begin{equation*}}
\def\eez{\end{equation*}}
\def\btez{\begin{eqnarray*}}
\def\etez{\end{eqnarray*}}
\def\sep{\mbox{,}}

\title{Phase-Field Model of Stressed Incoherent Solid-Solid Interfaces}

\author{J\'er\^ome Paret}
\email{jerome.paret@L2MP.u-3mrs.fr}
\affiliation{Laboratoire Mat\'eriaux et Micro\'electronique de Provence, CNRS UMR 6137, Facult\'e des Sciences de St J\'er\^ome, case 151, 13397 Marseille cedex 20, France}

\date{\today}

\begin{abstract}
We introduce a new phase-field model which allows for simulation of incoherent solid/solid transformations. Contrary to previous models which impose coherency at the interface, the zero shear-stress condition characteristic of incoherent solids is recovered in the limit of small interface thickness. For the sake of clarity, we limit ourselves to the case of stress-driven phase transitions between two elastic solids. However, since we use a variational formulation which has a clear thermodynamic interpretation, the extension of the model to the cases of chemical and/or thermal diffusion is straightforward. 
\end{abstract}

\maketitle

In recent years, phase-field models have become increasingly successful in simulating realistic microstructure formation during phase transitions. These models, which replace mathematically sharp interfaces by diffuse ones provide powerful numerical methods which avoid the tracking of the interface (free-boundary problem). The idea can be traced back to studies of phase transitions by Ginzburg and Landau \cite{Ginzburg50}, Cahn and Hilliard \cite{Cahn58}, Halperin et al. \cite{Halperin74} or Langer and Sekerka \cite{Langer75}, but it has received renewed interest because recent phase-field models make {\em quantitative} numerical simulations possible. The efforts of the community in this field can roughly be divided into two main directions : studies of solidification and dendritic growth on the one hand, coherent solid/solid transformations such as martensitic transformation on the other hand. Among the recent achievements within these two sub-fields one can cite the "thin-interface" limit of Karma and Rappel \cite{Karma98}, simulations of well developped 3D dendrites \cite{Karma00}, studies about grain boundary motion \cite{Lobkov01}, eutectic \cite{Elder00} or peritectic \cite{ShingLo01} growth and martensitic transformation \cite{Artemev01, Jin01}. Recently, Karma extended the model to include phases with asymmetric transport coefficients \cite{Karma01b} and some authors brought attention onto the problem of phase-field modelling of solid/fluid equilibria with elastic stresses. The latter studies were concerned with the so-called Asaro-Tiller-Grinfeld instability \cite{Muller99, Kassner01} or fracture phenomena \cite{Karma01a}. All of this is of course only a small sample of the existing litterature aimed at demonstrating the great variety of problems which can be tackled using this method. However, the interested reader will be able to find quite a large amount of references within these papers.

According to the thermodynamic description of solids given by Larch\'e and Cahn \cite{Larche78}, we can distinguish between three kinds of thermodynamic boundary conditions : solid/fluid, incoherent solid/solid or coherent solid/solid. The latter imposes continuity of the tractions at the interface whereas the other ones impose continuity of the normal traction and set the tangential component to zero. Only solid/fluid and coherent equilibria have been so far successfully modelled within the phase-field approach. Although solid/fluid and incoherent solid/solid equilibria appear to have the same kind of mechanical boundary conditions, their behavior become different when diffusion potentials are taken into account \cite{Larche78, Larche85}. Moreover, it was noticed by Kassner et al. \cite{Kassner01} that the direct generalization of their model would yield coherency at the interface. Thus, an "incoherent" model demands a treatment of its own. 

One of the main applications of such a model would be the correct description of intermetallic compounds formation by bulk isothermal diffusion, a phenomenon for which incoherent boundaries are rather the rule than the exception. These compounds can sometimes exhibit very irregular morphologies, as a consequence of a growth instability which is hard to explain if elastic stresses are not taken into account. This instability has been observed for example in Ni-Si \cite{Tu83} or Mo-Si binary couples. The problem is thus of practical importance since silicides are used as conducting layers in MOS (metal-oxyde-semiconductor) structures and the necessity to understand and control interfacial morphologies is obvious in this case.

In the following, we describe our model and derive its sharp-interface limit. This provides us with a Gibbs-Thomson condition for equilibrium at the interface. As a consequence of this relation, it will appear that a "Grinfeld-like" instability is possible at the interface between two uniaxially stressed incoherent solids.
 
\begin{figure}
\includegraphics[bb = 0 0 240 170]{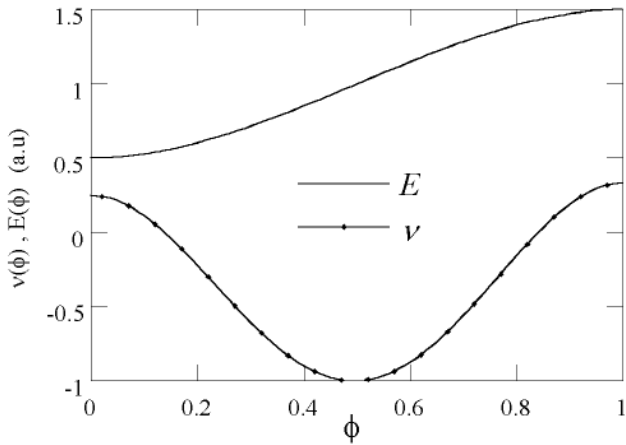}
\caption{Evolution of elastic coefficients $E$ and $\nu$ as a function of $\phi$  in the interfacial region $\phi \in [0\sep1]$.}
\label{fig_elastcoeff}
\end{figure}

We now introduce the model. Since it is supposed to yield quantitative modelling and most thermodynamic data are expressed at constant pressure, we choose to formulate the model with stresses as primary variables. Then, it is easy to derive free-energy contributions under non-hydrostatic stress conditions, using Maxwell relations to integrate from a state of constant pressure. We consider two solid phases $S_\alpha$ and $S_\beta$ which obeys Hooke's law with elastic coefficients $(E_{\alpha}\sep\nu_{\alpha})$ and $(E_{\beta}\sep\nu_{\beta})$ respectively. The underlying assumption of linear elasticity means that the strains $\varepsilon_{ij}$ have to be small, a condition which we will later make more quantitative. Note that this means that the applied stress $\sigma$ must remain much smaller than Young's modulus $E$. We introduce a phase-field $\phi$, along with a stress field $\sigma_{ij}$. We suppose that mechanical equilibrium is realized much faster than any other thermodynamic process which implies the relation $\partial_j \sigma_{ij}=0$. In this equation and all subsequent developments, Einstein convention is used. Subscripts preceded by a comma will denote spatial derivatives. We then postulate the following form for the free energy: 
\be
{\cal F}=\int dV \left\{\Gamma \left[\frac{d^2}{2}({\bf \nabla} \phi)^2 + 2g(\phi)\right]+ f_{el}(\phi,\{\sigma_{ij}\})\right\}\;\sep
\label{eqn:F}
\ee
\be
f_{el}(\phi,\{\sigma_{ij}\})=\frac{1+\nu(\phi)}{2E(\phi)}\sigma_{ij}\sigma_{ij}-\frac{\nu(\phi)}{2E(\phi)}\sigma_{kk}^2\;\sep
\label{eqn:elastF}
\ee
where $g(\phi)=\phi^2(1-\phi)^2$ is the usual double-well potential with minima at $\phi=0$ (phase $S_\alpha$) and $\phi=1$ (phase $S_\beta$) and $\Gamma=3\gamma/d$, $\gamma$ being the surface free-energy, which we take to be isotropic. The reference state for strain has been chosen as a state of zero stress and strain in both phases. The precise form of the functions $E(\phi)$ and $\nu(\phi)$ will only be needed later in the derivation. We plot them schematically in Fig.\ref{fig_elastcoeff}. Note that we have $E(0)=E_{\alpha}$, $E(1)=E_{\beta}$, $\nu(0)=\nu_{\alpha}$, $\nu(1)=\nu_{\beta}$, $E^{\prime}(0)=E^{\prime}(1)=0$, $\nu^{\prime}(0)=\nu^{\prime}(1)=0$.

We now have to specify how the fields evolve in time. Since we assumed mechanical equilibrium, ${\cal F}$ must be stationnary with respect to variations of $\sigma_{ij}$. As $\sigma_{ij}$ is constrained by the relation $\partial_j \sigma_{ij}=0$, some care is needed in writing down the equilibrium condition. Let us define the generalized strains $\varepsilon_{ij}$:
\be
\varepsilon_{ij} = \frac{\delta{\cal F}}{\delta\sigma_{ij}}=\frac{1+\nu(\phi)}{E(\phi)}\sigma_{ij}-\frac{\nu(\phi)}{E(\phi)}\sigma_{kk}\delta_{ij}\;.
\label{eqn:Hooke}
\ee
Then, it can be shown that the stationnarity condition is:
\be
\varepsilon_{ij,kk}+\varepsilon_{kk,ij}-\varepsilon_{ik,jk}-\varepsilon_{jk,ik}=0\;\sep \label{eqn:compat}
\ee
which is just St Venant's compatibility equation expressing the fact that the strain tensor $\varepsilon_{ij}$ is the symetric part of the gradient of a displacement field.

For the phase-field $\phi$, we impose non-conserved dynamics (Model A of the classification by Halperin et al. \cite{Halperin74}). We thus have the equation: 
\be
\frac{\partial \phi}{\partial t}= -R\frac{\delta {\cal F}}{\delta \phi}
\ee
where $R=1/(3\tilde{k}\epsilon)$ is a phenomenological rate constant. Expanding and casting the last equation into nondimensional form, we obtain:
\be
\frac{\partial \phi}{\partial t}=\nabla^2\phi-\frac{2g^{\prime}}{\epsilon^2}-\frac{1}{3\epsilon}\left\{{\left[\frac{1+\nu}{2E}\right]}^{\prime}\sigma_{ij}\sigma_{ij}-{\left[\frac{\nu}{2E}\right]}^{\prime}\sigma_{kk}^2\right\}
\label{eqn:kinetic}
\ee
with $\epsilon=d/l_G$ and $l_G=\gamma E/\sigma^2$ is the Griffith's length ($E$ average Young's modulus, $\sigma$ applied stress). Since the so-called "sharp-interface limit", which allows to recover the usual Gibbs-Thomson equation, is a matched asymptotic expansion with $\epsilon$ as a small parameter, the interface thickness must be much smaller than the Griffith's length in order for the phase-field model to be useful. This sets a limit on the range of applied stresses which we can be simulated within this framework. Even if other phenomena such as chemical or thermal diffusion are present and set a natural length scale, it is always possible to find a range of external stresses small enough to keep the elastic term in (\ref{eqn:kinetic}) much smaller than the double-well potential term, a condition which is necessary in order to obtain meaningful asymptotics. Indeed, the extension to the case of a diffusive transformation is straightforward as long as no cross-effects (such as thermal expansion or compositionally-induced stresses) are present: we just need to add the elastic part (\ref{eqn:elastF}) to free-energies used in "diffusive" models. Since there is no bulk coupling between stress and temperature or concentration fields, they can be treated as independent variables and the analysis leading to thin or sharp-interface limits is not modified. 

From now on, all length scales will be measured in units of $l_G$, which amounts to replacing $d$ by $\epsilon$ in (\ref{eqn:F}). Moreover, we will limit the following discussion to the two-dimensional case, the three-dimensional case being just a straightforward generalization of the latter.

We now perform the asymptotic expansion. In the outer domain, i.e. the domain in which gradients of $\phi$ vanish, the phase-field equation (\ref{eqn:kinetic}) gives the condition $g^{\prime}(\phi)=0$ at lowest order in $\epsilon$. This equation yields three solutions, $\phi=0\sep\;1$ or 1/2, the latter being unstable. Since $g^{\prime}$, $E^{\prime}$ and $\nu^{\prime}$ vanish when $\phi=0\;\mbox{or}\;1$, these two solutions are valid at all orders of the expansion. The outer equations for the elastic fields then read:
\be
(1+\nu_p)\sigma_{ij,kk}+\sigma_{kk,ij}=0\;\;\sep\;\;p=(\alpha\,\sep\,\beta)\;\;\sep
\ee
which are just the usual Beltrami-Mitchell equations for each phase.

Let us now consider the inner equations, which solutions will allow to match the outer solutions at the interface. We switch to a curvilinear coordinates system comoving with the interface. We choose the $r$ axis to be perpendicular to the interface (i.e. parallel to $\vec{\nabla\phi}$) and oriented from phase $S_\beta$ to phase $S_\alpha$. The other coordinate $s$ is taken along the tangent to the curve $\phi=1/2$. We then stretch the $r$ variable, introducing $\rho=r/\epsilon$. Using capital letters for the inner fields, we obtain at leading order:
\be
\partial^2_{\rho}\Phi^{(0)}=2g^{\prime}(\Phi^{(0)})
\label{eqn:innerPhi0}
\ee
for the phase-field and:
\be
\partial_{\rho}\Sigma^{(0)}_{\rho\rho}=0\;\;\sep\;\;\partial_{\rho}\Sigma^{(0)}_{\rho s}=0\;\;\sep\;\;\partial^2_{\rho}E^{(0)}_{ss}=0
\label{eqn:innerElast0}
\ee
for the elastic fields. From equation (\ref{eqn:innerElast0}), we obtain that $\Sigma^{(0)}_{\rho\rho}$, $\Sigma^{(0)}_{\rho s}$ and $E^{(0)}_{ss}$ do not depend on $\rho$ within the inner region, a conclusion identical to the results by Kassner et al. \cite{Kassner01}.

We now give the detailed form of functions $\nu(\phi)$ and $E(\phi)$. We define the function $h(\phi)=\phi^2(3-2\phi)$. We then have :
\bte
& E(\phi)=E_{\alpha}+(E_{\beta}-E_{\alpha})h(\phi) & \\
& 1+\nu(\phi)={(aE(\phi)-b)}^2 &\sep\label{eqn:poisson}
\ete
where $a$ and $b$ are defined according to:
\btez
& (E_{\beta}-E_{\alpha})\,a = \sqrt{1+\nu_{\alpha}}+\sqrt{1+\nu_{\beta}} & \\
& (E_{\beta}-E_{\alpha})\,b = E_{\beta}\sqrt{1+\nu_{\alpha}}+E_{\alpha}\sqrt{1+\nu_{\beta}} &\mbox{.}
\etez

With this choice of functions, it is easily verified that we have $\nu(0)=\nu_{\alpha}$, $\nu(1)=\nu_{\beta}$, $E(0)=E_{\alpha}$, $E(1)=E_{\beta}$, that the derivatives $\nu^{\prime}$ and $E^{\prime}$ are zero when $\phi=0$ or $1$ and that there exists $\phi_c \in [0\sep 1]$ such that $1+\nu(\phi_c)=\nu^{\prime}(\phi_c)=0$. 
Since we have:
\be
E^{(0)}_{\rho s}=\frac{1+\nu(\Phi^{(0)})}{E(\Phi^{(0)})}\Sigma^{(0)}_{\rho s}=T(s,t)\;\sep
\ee
and $1+\nu(\phi)$ vanishes somewhere in the inner region, $T(s,t)$ and thus $\Sigma^{(0)}_{\rho s}$ must be zero. We now match the inner and outer fields, using the matching condition
\be
\psi(r,s,t)|_{r=0^{\pm}}=\Psi(\rho,s,t)|_{\rho\to\pm\infty}\;\sep
\ee 
valid for any physical field $\psi$. Equation (\ref{eqn:innerPhi0}) yields the classical $\Phi^{(0)}=(1-\tanh\rho)/2$ solution which obviously matches the outer solutions $\phi=0$ and $\phi=1$. For the stress field, we obtain:
\be
{\sigma^{(\beta)}_{rr}}|_{r=0^{-}}=\sigma^{(\alpha)}_{rr}|_{r=0^{+}}\;\;\sep\;\;{\sigma^{(\beta)}_{rs}}|_{r=0^{-}}=\sigma^{(\alpha)}_{rs}|_{r=0^{+}}=0\;\sep
\ee
which are the mechanical equilibrium conditions for incoherent interfaces.

At first order in the expansion for $\Phi$, we obtain an equation of the form $L\Phi^{(1)}=V(\Phi^{(0)},\Sigma^{(0)}_{ij})$ where we defined the following self-adjoint operator: $L\equiv\partial^2_{\rho}-2g^{\prime\prime}(\Phi^{(0)})$. In order for a non-trivial solution to exist, the right-hand side must be orthogonal to the zero-mode $\partial_{\rho}\Phi^{(0)}$. This condition is called the "solvability condition" and reads, using $\Sigma^{(0)}_{\rho s}=0$ and evaluating $\phi$-derivatives at $\phi=\Phi^{(0)}$:
\begin{widetext}
\be
0=\int_{-\infty}^{+\infty}d\rho\partial_\rho \Phi^{(0)}\left\{-3\partial_\rho \Phi^{(0)}\left(\tilde{k}v+\gamma\kappa\right)+\frac{d}{d\phi}\left[\frac{1+\nu}{2E}\right]\left({\Sigma^{(0)}_{rr}}^2+{\Sigma^{(0)}_{ss}}^2\right)-\frac{d}{d\phi}\left[\frac{\nu}{2E}\right]{\left(\Sigma^{(0)}_{rr}+\Sigma^{(0)}_{ss}\right)}^2\right\}\;\mbox{.}
\label{eqn:solvability}
\ee
\end{widetext}

We finally note that $E$ is a monotonous function of $\phi$, which allows for a change of variables in the interval $\phi \in [0\sep 1]$. Using equations (\ref{eqn:Hooke}), (\ref{eqn:innerElast0}) and (\ref{eqn:poisson}) to express $\Sigma^{(0)}_{ss}$ as a function of $E_{ss}(\Phi^{(0)})$, performing the change of variables and computing the integral in (\ref{eqn:solvability}), we finally obtain for the local velocity of the interface $v$:
\be
v=-\frac{1}{\tilde{k}}\left\{\frac{E_{\beta}-E_{\alpha}}{2E_{\beta}^2}{(\sigma_{ss}^{(\beta)}-\sigma_{rr}^{(\beta}))}^2+\gamma\kappa\right\}\;\sep
\label{eqn:Gibbs}
\ee 
which is the Gibbs-Thomson equation for our model. In doing so, we assumed $\Sigma_{\rho\rho}^{(0)}=0$ for simplicity. This choice is natural since we chose the equilibrium state with zero stress as the reference state for strains.

When performing the above analysis in two-dimensional space, we implicitly assumed that the elastic fields corresponded to a state of plane stress ($\sigma_{zz}=0$). The plane strain case ($\varepsilon_{zz}=0$) is easily obtained from equation (\ref{eqn:Gibbs}), replacing $E$ by $E/(1-\nu^2)$. In the latter case, we could also assume that phase $S_\alpha$ describes a fluid, which gives $E_{\alpha}=0$. Then, we recover exactly the same Gibbs-Thomson equation as was obtained by Kassner et al. in their study of Grinfeld instability \cite{Kassner01}. This in turn qualitatively shows that a "Grinfeld-like" instability is also possible at an incoherent solid/solid interface.

In the present state, our model is restricted to pure "stress-transitions" whereas an incoherent phase transformation almost always involves mass transport across the interface. This means that the model should at least be extended to the case of chemical diffusion, which is quite straightforward as explained earlier. Going further toward realistic systems, it would be interesting to incorporate stresses induced by compositional inhomogeneities into the picture. Such an analysis has been performed by Gurtin \cite{Gurtin96} for the case of coherent equilibrium, using a phase-field model with a conserved order parameter (Cahn-Hilliard equation) and a different thermodynamic approach. In the present case, the extension is not conceptually difficult since the variational formulation and its thermodynamic interpretation allow for an easy translation of Larch\'e and Cahn analysis \cite{Larche85} into "phase-field language". However, it can not be decided from such qualitative arguments whether a zero-th order expansion for the inner elastic fields would be sufficient to recover the correct Gibbs-Thomson equation.

To summarize, we have developped a new phase-field model aimed at simulating stressed incoherent solid/solid interfaces. In the limit where Young's modulus of one of the phase goes to zero, the Gibbs-Thomson equation for solid/fluid equilibrium with elastic stresses is recovered \cite{Kassner01}. However, contrary to the model by Kassner et al., the present model is formulated with stress as a primary field which makes it easier to link to thermodynamic data for quantitative simulations. Moreover, such a formulation is easier to simulate when applied to systems where stress is imposed at external boundaries, the Neumann boundary conditions for the Navier equations becoming Dirichlet conditions for the Beltrami-Mitchell equations. The model lends itself to easy generalization to more complex situations (thermal or chemical difusion, thermo-elastic materials, compositionally-induced stresses ...). The specific case of incoherent equilibrium with diffusion and chemical self-stresses will be the object of a future publication.
  
\begin{acknowledgments}
This work was supported by Centre National de la Recherche Scientifique and Universit\'e Aix-Marseille III. We are grateful to B. Billia, F. Celestini, J.M. Debierre, P. Gas and R. Gu\'erin for stimulating discussions and support. K. Kassner and A. Karma are acknowledged for providing useful comments on early ideas about this study.
\end{acknowledgments}


\begin{thebibliography}{19}
\expandafter\ifx\csname natexlab\endcsname\relax\def\natexlab#1{#1}\fi
\expandafter\ifx\csname bibnamefont\endcsname\relax
  \def\bibnamefont#1{#1}\fi
\expandafter\ifx\csname bibfnamefont\endcsname\relax
  \def\bibfnamefont#1{#1}\fi
\expandafter\ifx\csname citenamefont\endcsname\relax
  \def\citenamefont#1{#1}\fi
\expandafter\ifx\csname url\endcsname\relax
  \def\url#1{\texttt{#1}}\fi
\expandafter\ifx\csname urlprefix\endcsname\relax\def\urlprefix{URL }\fi
\providecommand{\bibinfo}[2]{#2}
\providecommand{\eprint}[2][]{\url{#2}}

\bibitem[{\citenamefont{Ginzburg and Landau}(1950)}]{Ginzburg50}
\bibinfo{author}{\bibfnamefont{V.}~\bibnamefont{Ginzburg}} \bibnamefont{and}
  \bibinfo{author}{\bibfnamefont{L.~D.} \bibnamefont{Landau}},
  \bibinfo{journal}{JETP} \textbf{\bibinfo{volume}{20}}, \bibinfo{pages}{1064}
  (\bibinfo{year}{1950}).

\bibitem[{\citenamefont{Cahn and Hilliard}(1958)}]{Cahn58}
\bibinfo{author}{\bibfnamefont{J.}~\bibnamefont{Cahn}} \bibnamefont{and}
  \bibinfo{author}{\bibfnamefont{J.~E.} \bibnamefont{Hilliard}},
  \bibinfo{journal}{J. Chem. Phys.} \textbf{\bibinfo{volume}{28}},
  \bibinfo{pages}{258} (\bibinfo{year}{1958}).

\bibitem[{\citenamefont{Halperin et~al.}(1974)\citenamefont{Halperin,
  Hohenberg, and Ma}}]{Halperin74}
\bibinfo{author}{\bibfnamefont{B.}~\bibnamefont{Halperin}},
  \bibinfo{author}{\bibfnamefont{P.~C.} \bibnamefont{Hohenberg}},
  \bibnamefont{and} \bibinfo{author}{\bibfnamefont{S.~K.} \bibnamefont{Ma}},
  \bibinfo{journal}{Phys. Rev. B} \textbf{\bibinfo{volume}{10}},
  \bibinfo{pages}{139} (\bibinfo{year}{1974}).

\bibitem[{\citenamefont{Langer and Sekerka}(1975)}]{Langer75}
\bibinfo{author}{\bibfnamefont{J.~S.} \bibnamefont{Langer}} \bibnamefont{and}
  \bibinfo{author}{\bibfnamefont{R.~F.} \bibnamefont{Sekerka}},
  \bibinfo{journal}{Acta metall.} \textbf{\bibinfo{volume}{23}},
  \bibinfo{pages}{1225} (\bibinfo{year}{1975}).

\bibitem[{\citenamefont{Karma and Rappel}(1998)}]{Karma98}
\bibinfo{author}{\bibfnamefont{A.}~\bibnamefont{Karma}} \bibnamefont{and}
  \bibinfo{author}{\bibfnamefont{W.~J.} \bibnamefont{Rappel}},
  \bibinfo{journal}{Phys. Rev. E} \textbf{\bibinfo{volume}{57}},
  \bibinfo{pages}{4323} (\bibinfo{year}{1998}).

\bibitem[{\citenamefont{Karma et~al.}(2000)\citenamefont{Karma, Lee, and
  Plapp}}]{Karma00}
\bibinfo{author}{\bibfnamefont{A.}~\bibnamefont{Karma}},
  \bibinfo{author}{\bibfnamefont{Y.~H.} \bibnamefont{Lee}}, \bibnamefont{and}
  \bibinfo{author}{\bibfnamefont{M.}~\bibnamefont{Plapp}},
  \bibinfo{journal}{Phys. Rev. E} \textbf{\bibinfo{volume}{61}},
  \bibinfo{pages}{3996} (\bibinfo{year}{2000}).

\bibitem[{\citenamefont{Lobkovsky and Warren}(2001)}]{Lobkov01}
\bibinfo{author}{\bibfnamefont{A.~E.} \bibnamefont{Lobkovsky}}
  \bibnamefont{and} \bibinfo{author}{\bibfnamefont{J.~A.}
  \bibnamefont{Warren}}, \bibinfo{journal}{Phys. Rev. E}
  \textbf{\bibinfo{volume}{63}}, \bibinfo{pages}{051605}
  (\bibinfo{year}{2001}).

\bibitem[{\citenamefont{Drolet et~al.}(2000)\citenamefont{Drolet, Elder, Grant,
  and Kosterlitz}}]{Elder00}
\bibinfo{author}{\bibfnamefont{F.}~\bibnamefont{Drolet}},
  \bibinfo{author}{\bibfnamefont{K.~R.} \bibnamefont{Elder}},
  \bibinfo{author}{\bibfnamefont{M.}~\bibnamefont{Grant}}, \bibnamefont{and}
  \bibinfo{author}{\bibfnamefont{J.~M.} \bibnamefont{Kosterlitz}},
  \bibinfo{journal}{Phys. Rev. E} \textbf{\bibinfo{volume}{61}},
  \bibinfo{pages}{6705} (\bibinfo{year}{2000}).

\bibitem[{\citenamefont{Shing-Lo et~al.}(2001)\citenamefont{Shing-Lo, Karma,
  and Plapp}}]{ShingLo01}
\bibinfo{author}{\bibfnamefont{T.}~\bibnamefont{Shing-Lo}},
  \bibinfo{author}{\bibfnamefont{A.}~\bibnamefont{Karma}}, \bibnamefont{and}
  \bibinfo{author}{\bibfnamefont{M.}~\bibnamefont{Plapp}},
  \bibinfo{journal}{Phys. Rev. E} \textbf{\bibinfo{volume}{63}},
  \bibinfo{pages}{031504} (\bibinfo{year}{2001}).

\bibitem[{\citenamefont{Artemev et~al.}(2001)\citenamefont{Artemev, Jin, and
  Khachaturyan}}]{Artemev01}
\bibinfo{author}{\bibfnamefont{A.}~\bibnamefont{Artemev}},
  \bibinfo{author}{\bibfnamefont{Y.~M.} \bibnamefont{Jin}}, \bibnamefont{and}
  \bibinfo{author}{\bibfnamefont{A.~G.} \bibnamefont{Khachaturyan}},
  \bibinfo{journal}{Acta Mater.} \textbf{\bibinfo{volume}{49}},
  \bibinfo{pages}{1165} (\bibinfo{year}{2001}).

\bibitem[{\citenamefont{Jin et~al.}(2001)\citenamefont{Jin, Artemiev, and
  Khachaturyan}}]{Jin01}
\bibinfo{author}{\bibfnamefont{Y.~M.} \bibnamefont{Jin}},
  \bibinfo{author}{\bibfnamefont{A.}~\bibnamefont{Artemiev}}, \bibnamefont{and}
  \bibinfo{author}{\bibfnamefont{A.~G.} \bibnamefont{Khachaturyan}},
  \bibinfo{journal}{Acta Mater.} \textbf{\bibinfo{volume}{49}},
  \bibinfo{pages}{2309} (\bibinfo{year}{2001}).

\bibitem[{\citenamefont{Karma}(2001)}]{Karma01b}
\bibinfo{author}{\bibfnamefont{A.}~\bibnamefont{Karma}} (\bibinfo{year}{2001}),
  \eprint{cond-mat/0103289}.

\bibitem[{\citenamefont{M{\"u}ller and Grant}(1999)}]{Muller99}
\bibinfo{author}{\bibfnamefont{J.}~\bibnamefont{M{\"u}ller}} \bibnamefont{and}
  \bibinfo{author}{\bibfnamefont{M.}~\bibnamefont{Grant}},
  \bibinfo{journal}{Phys. Rev. Lett.} \textbf{\bibinfo{volume}{82}},
  \bibinfo{pages}{1736} (\bibinfo{year}{1999}).

\bibitem[{\citenamefont{Kassner et~al.}(2001)\citenamefont{Kassner, Misbah,
  M{\"u}ller, Kappey, and Kohlert}}]{Kassner01}
\bibinfo{author}{\bibfnamefont{K.}~\bibnamefont{Kassner}},
  \bibinfo{author}{\bibfnamefont{C.}~\bibnamefont{Misbah}},
  \bibinfo{author}{\bibfnamefont{J.}~\bibnamefont{M{\"u}ller}},
  \bibinfo{author}{\bibfnamefont{J.}~\bibnamefont{Kappey}}, \bibnamefont{and}
  \bibinfo{author}{\bibfnamefont{P.}~\bibnamefont{Kohlert}},
  \bibinfo{journal}{Phys. Rev. E} \textbf{\bibinfo{volume}{63}},
  \bibinfo{pages}{036117} (\bibinfo{year}{2001}).

\bibitem[{\citenamefont{Karma et~al.}(2001)\citenamefont{Karma, Kessler, and
  Levine}}]{Karma01a}
\bibinfo{author}{\bibfnamefont{A.}~\bibnamefont{Karma}},
  \bibinfo{author}{\bibfnamefont{D.~A.} \bibnamefont{Kessler}},
  \bibnamefont{and} \bibinfo{author}{\bibfnamefont{H.}~\bibnamefont{Levine}},
  \bibinfo{journal}{Phys. Rev. Lett.} \textbf{\bibinfo{volume}{87}},
  \bibinfo{pages}{045501} (\bibinfo{year}{2001}).

\bibitem[{\citenamefont{Larch\'e and Cahn}(1978)}]{Larche78}
\bibinfo{author}{\bibfnamefont{F.}~\bibnamefont{Larch\'e}} \bibnamefont{and}
  \bibinfo{author}{\bibfnamefont{J.~W.} \bibnamefont{Cahn}},
  \bibinfo{journal}{Acta metall.} \textbf{\bibinfo{volume}{26}},
  \bibinfo{pages}{1579} (\bibinfo{year}{1978}).

\bibitem[{\citenamefont{Larch\'e and Cahn}(1985)}]{Larche85}
\bibinfo{author}{\bibfnamefont{F.}~\bibnamefont{Larch\'e}} \bibnamefont{and}
  \bibinfo{author}{\bibfnamefont{J.~W.} \bibnamefont{Cahn}},
  \bibinfo{journal}{Acta metall.} \textbf{\bibinfo{volume}{33}},
  \bibinfo{pages}{331} (\bibinfo{year}{1985}).

\bibitem[{\citenamefont{Tu et~al.}(1983)\citenamefont{Tu, Ottaviani,
  G{\"o}sele, and F{\"o}ll}}]{Tu83}
\bibinfo{author}{\bibfnamefont{K.~N.} \bibnamefont{Tu}},
  \bibinfo{author}{\bibfnamefont{G.}~\bibnamefont{Ottaviani}},
  \bibinfo{author}{\bibfnamefont{U.}~\bibnamefont{G{\"o}sele}},
  \bibnamefont{and} \bibinfo{author}{\bibfnamefont{H.}~\bibnamefont{F{\"o}ll}},
  \bibinfo{journal}{J. Appl. Phys.} \textbf{\bibinfo{volume}{54}},
  \bibinfo{pages}{758} (\bibinfo{year}{1983}).

\bibitem[{\citenamefont{Gurtin}(1996)}]{Gurtin96}
\bibinfo{author}{\bibfnamefont{M.~E.} \bibnamefont{Gurtin}},
  \bibinfo{journal}{Physica D} \textbf{\bibinfo{volume}{92}},
  \bibinfo{pages}{178} (\bibinfo{year}{1996}).

\end{thebibliography}
\end{document}